\documentclass[12pt]{article}
\usepackage{lingmacros}
\usepackage[english]{babel}
\usepackage{graphicx}

\large \baselineskip = 20 true pt

\begin{document}
\begin{center}
{\Large \bf Implementation of Simplex Channels in the Blom's Keys Pre-Distribution Scheme. } \vspace{0.5cm}
\end{center}

\begin{center}
S.V. Belim, S.Yu. Belim \\
Dostoevsky Omsk State University, Omsk, Russia
 \vspace{0.5cm}
\end{center}

\begin{center}
{\bf Abstract}
\end{center}

{\small
In article the modification of the Blom's keys preliminary distribution scheme,  considering 
the direction of information streams is suggested. For this modification it is necessary to use
function from three variables. Function of formation of key materials will be the asymmetrical. 
The exponential form of this function which does not increase the volume of key materials 
is suggested.
}


\section{Introduction}

Keys preliminary distribution schemes are used for the organization of users interaction on 
secure channels in communication network. The main idea consists that each user receives a set 
of key materials on the basis of which he can calculate the common key of enciphering with any
other user of system, using open information.

The Blom's  keys preliminary distribution scheme is based on interaction model "everyone with
everyone". However such approach can be used only in networks with the entrusted users. 
In the modern global networks the approach in which there are restrictions for the privileges 
of users is used. There can be restrictions as on actions of the user in all network, 
and on access to resources. There are several ways for solution these tasks. One of approaches
received the name ID-based cryptography \cite{b1}.

Now problems of ID-based cryptography are solved with the enciphering asymmetric algorithms. 
The main idea consists in development of key information on the basis of identifiers \cite{b1} 
or user attributes \cite{b2}. Recently ID-based cryptography develops in the direction 
of the approaches allowing contacting any subscriber of network only on the basis of open
information. Each user of network has to receive some certificate from certification center. 
In many articles problems of sending the ciphered messages and the digital signature by means 
of cryptographic algorithms are solved with an open key. The main lack of this approach is the
common problem of asymmetric cryptography use - larger complexity of calculations. 
It brings to operating time delays. Therefore, the problem of development of ID-based 
cryptography fast algorithms is relevant.

It is also necessary to consider criteria to which the attributes based cryptosystem, formulated 
in article \cite{b3}, has to satisfy:

K1: Data confidentiality.

Data shall be encrypted by the owner before their sending. The unauthorized user cannot learn about data which were encrypted.

K2: Detail access monitoring.

In users group the system provides various access rules for each certain participant of group.
Thus, users from one group can have various access rules to data.

K3: Scalability.

The number of the registered users should not influence systems efficiency.

K4: Monitoring of actions.

Transfer the authorized user secret key attributes to other persons is inadmissible.

K5: Response of the user's rights.

If the user quits the system, then the system can recall the rights of this user. The user, 
whose rights were recall, will not be able to get access to data any more.

K6: Impossibility of conspiracy.

Users cannot unite the attributes to decipher data. Each attribute is bound to a polynomial 
or a random number. Thus, users cannot enter into a collision with each other. Also distribution 
of keys with restriction for interaction can be realized by means of hash functions \cite{b4}.

Other approach is based on modification of keys preliminary distribution schemes. In article
\cite{b5} the modification of the Blom's scheme allowing realizing the ban on certain information
channels between users is suggested. In articles \cite{b6,b7} the similar task is solved on the
basis of the KDP scheme keys preliminary distribution. In both cases the global exchange 
by information between two users is forbidden or allowed. In actual systems it is necessary 
to solve a access control problem for different type resources. Thus, the problem of keys
preliminary distribution schemes modification, considering the direction of informational 
streams, is relevant.

The purpose of this article is development of the keys preliminary distribution schemes for system taking into account types of accesses.

\section{Problem definition}

Let's consider system with two basic types of access: read ($r$) and write ($w$). At 
initialization of the information channel it is necessary to watch in what direction data will 
be transferred. Thus, all channels have to be simplex. Let's enter the variable s for direction 
of data transmission. If the user $u$ sends a request for opening the channel on reading, 
that is obtaining information, then $s=-1$. If the user opens the channel on writing, that is
information transfer, then $s=1$.

In the keys preliminary distribution scheme to each user via a secure channel key materials are
given. On the basis of key materials with use open information the user calculates an encrypting
key. In case of simplex channels for couple of users $u_i$ and $u_j$ two keys have to be formed:
$k_{ij}$ -- key for information transfer from the user $u_i$ to the user $u_j$, $k_{ji}$ -- key 
for information transfer from the user $u_j$ to the user $u_i$. In case of inquiry of the user 
of $u_i$ for reading information to the user of $u_j$ or inquiry of the user of $u_j$ to the user
of $u_i$ for record $k_{ji}$ key is used. In case of inquiry of the user of $u_j$ for reading
information to the user of $u_i$ or inquiry of the user of $u_i$ to the user of $u_j$ for record
$k_{ij}$ key is used. Let's demand that the forbidden channels of communication had a zero pair
key, and legal -- nonzero. We will call the keys preliminary distribution scheme meeting these
conditions simplex.

\section{Algorithm for keys preliminary distribution}

We realize the scheme of preliminary distribution of keys taking into account simplex channels 
on the basis of the well-known Blom's scheme. In the Blom's keys preliminary distribution scheme 
on the server the symmetric polynomial from two variables $f(x,y)$ over the field $Z_p$  is
generated. Some number $r_i$ is compared to each user $u_i$. Further for each user polynomials 
from one variable are formed $g_i(x)=f(x,r_i)$. These polynomials are transferred via secure
channels to users and saved in a secret. If necessary to develop the common key with the user 
$u_j$ the user $u_i$ takes the $r_j$ element from open base and calculates value 
$k_{ij}=g_i(r_j)$. The same way the user of $u_j$ enters, calculating $k_{ji}=g_j(r_i)$. 
Symmetry of a polynomial $f(x,y)$ provides equality $k_{ij}=k_{ji}$.

For implementation of the simplex keys preliminary distribution scheme it is necessary 
to consider the directions of informational streams. For accounting of the direction of streams 
we will generate on the server a polynomial from three variables $F(x,y,s)$ over the field $Z_p$.
The variable s can accept two values (+1 or-1) and defines the direction of information stream. 
To each user $u_i$ it is comparable some number $r_i$ which is sited in open form and it is
protected from changes. For each user keys distribution server forms a polynomial 
$G_i(x,s)=F(x,r_i,s)$ also transfers it via a secure channel. On the basis of the function 
$G_i(x,s)$ and numbers $r_i$ for the allowed information channel  $k_{ij}\neq0$, and for the
forbidden channels $k_{ij}=0$ is generated. Both participants of exchange have to develop 
an identical key on the basis of secret key materials and open information about each other.

Let's consider two types of user's interaction at information transfer. First, we will stop 
on a case in which the user of $u_i$ initiates obtaining information from the user of $u_j$, 
that is makes to it an inquiry on reading. The user $u_j$ sends information, beforehand having
encrypts it a key
\[
k_{ij}=G_j (r_i,1)=F(r_i,r_j,1).
\]

The parameter $s=1$, as for user $u_j$ the informational stream is proceeding. The user $u_i$
receives the message and decrypts it the same key, calculated on the basis of the key materials:
\[
k_ji=G_i (r_j,-1)=F(r_j,r_i,-1).
\]
The parameter $s= -1$, as for $u_i$ the informational stream is entering.

The second case consists that the user $u_i$ initiates sending information to the user $u_j$,
making to it an inquiry on record. The user $u_i$ sends information, beforehand having encrypted 
it a key
\[
k_ij=G_i (r_j,1)=F(r_j,r_i,1).
\]
The parameter $s=1$, as for $u_i$ the informational stream is proceeding. The user $u_j$ receives
the message and decrypts it the same key, calculated on the basis of the key materials:
\[
k_ij=G_j (r_i,-1)=F(r_j,r_i,-1).
\]
The parameter $s= -1$, as for $u_j$ the informational stream is entering.
From consideration of these two cases three requirements imposed on function $F(x,y,s)$:
\[
F(x,y,1)\neq F(y,x,1),\ \ 
F(x,y,-1)\neq F(y,x,-1),\ \ 
F(x,y,1)=F(y,x,-1).
\]
The first two requirements come down to asymmetry of function $F(x,y,s)$ to shift of the first 
two arguments at any value of the third argument. For cryptographic firmness it is also necessary
to demand that function $F(x,y,s)$ evaluation was impossible on the basis of the known value
$F(y,x,s)$.
	
Let's set function F(x,y,s) in a form:
\[
F(x,y,1)=x^{h(y)},\ \ F(x,y,-1)=y^{h(x)}.
\]
$h(x)$ is some polynomial from one variable. This function meets all three requirements, which 
are listed above. The cryptographic firmness of this function is provided with complexity 
of integer logarithm problems and calculation of a root in terminating fields.

Let's consider the keys distribution scheme on the basis of this function. Key materials are
transferred to each user. The user calculates function from two variables:
\[
G_i (x,1)=F(x,r_i,1)=x^{h(r_i)},
G_i (x,-1)=F(x,r_i,-1)=r_i^{h(x)}.
\]
For calculation the function values $G_i(x,1)$ the user of $u_i$ needs to know one number 
$h_i=h(r_i)$ which the server transfers it via a secure channel.

Let's consider more composite case of calculation $G_i(x,-1)$. Let $h(x)$ is polynom:
\[
h(x)=a_m x^m+a_{m-1} x^{m-1}+...+a_1 x+a_0.
\]
Then value of function:
\[
G_i (x,1)=F(x,r_i,1)=x^{h(r_i)}=(b_m^{(i)})^{z_m}...(b_1^{(i)})^{z_1} b_0^{(i)},
\]
there
\[
b_k^{(i)}=r_i^{a_k},\ \ z_k=x^k\ \ \ (k=1,...,m).
\]
Thus, the server as key materials sends to the user $u_i$ via a secure channel the vector:
\[
g_i=\left(h_i,b_m^{(i)},b_{m-1}^{(i)},...,b_1^{(i)},b_0^{(i)}\right).
\]
These data are enough for calculation the values of function. The problem of coefficients
determination for polynomial $h(x)$ from vector coordinates $g_i$ comes down to a discrete
logarithm.

This scheme slightly increases the volume of key materials. Only one number is added to the
traditional Blom's scheme.

\section*{Conclusion}

Modification of the Blom's scheme allows organizing messages exchange on simplex channels. 
For this purpose it is necessary to refuse the idea on use of symmetric polynomials. 
Such approach is caused by the fact that exchange of information becomes the asymmetrical.
Calculation of a key demands use the operation of an involution. It is characteristic for all
algorithms ID-based cryptography. However the suggested scheme allows controlling the directions 
of informational streams that is its distinctiveness.


\begin{thebibliography}{00}

\bibitem{b1}
Shamir A. Identity-Based Cryptosystems and Signature Schemes Advances in Cryptology. Proceedings of CRYPTO 84, Lecture Notes in Computer Science, 7, pp. 47--53 (1984).
\bibitem{b2}
Boneh D. and Franklin M. Identity-based encryption from the weil pairing. CRYPTO '01: 
Proceedings of the 21st Annual International Cryptology Conference on Advances in Cryptology, 
pp. 213--229 (2001).
\bibitem{b3}
Lee C.-C.,  Chung P.-S. and Hwang M.-S. A Survey on Attribute-based Encryption Schemes of Access Control in Cloud Environments. Int.J. of Network Security, 15, 4, pp.231--240 (2013).
\bibitem{b4}
Belim S.V. and Bogachenko N.F. Distribution of Cryptographic Keys in Systems with a Hierarchy of Objects.  Automatic Control and Computer Sciences, 50, 8, pp.773--776. (2016)
\bibitem{b5}
Belim S.V., Belim S.Yu, Polyakov S.Yu. The Implementation of Discretionary Access Separation Using a Modified Blom's Scheme of Key Distribution. Information Security Problems. Computer Systems, 3, 
pp. 72--76. (2015)
\bibitem{b6}
Belim S.V. and Belim S.Yu. KDP Scheme of Preliminary Key Distribution in Discretionary Security Policy.  Automatic Control and Computer Sciences, 50, 8, pp.777-786. (2016)
\bibitem{b7}
Belim S.V. and Belim S.Yu. The VPN Implementation on Base of the KDP-Scheme. CEUR Workshop Proceedings 1732. URL: http://ceur-ws.org/Vol-1732/paper3.pdf. (2016)
\end{thebibliography}
\end{document}